\documentclass[notoc]{JHEP3}

\usepackage{axodraw}

\newcommand{\ra}{\rangle}
\newcommand{\la}{\langle}
\renewcommand{\P}{\mathcal P}
\newcommand{\Phat}{\widehat P}
\newcommand{\N}{\mathcal N}

\title{A direct proof of the CSW rules}
\author{Kasper Risager,\\
Niels Bohr Institute, University of Copenhagen,
  Blegdamsvej 17, DK--2100 Copenhagen \O, Denmark
\\ Email: \email{risager@nbi.dk}
}
\abstract{
  Using recursion methods similar to those of Britto, Cachazo, Feng
  and Witten (BCFW) a direct proof of the CSW rules for computing
  tree-level gluon amplitudes is given. }
\preprint{hep-th/0508206}

\begin{document}

\section{Introduction}

Over the last 18 months there has been significant improvement in the
methods for calculating scattering amplitudes, primarily in gauge
theories but also in more general quantum field theory settings. The
progress was initiated by Witten's proposal \cite{Witten:2003nn}
(based partly on an earlier insight by Nair \cite{Nair:1988bq}) that
$\N=4$ Super Yang--Mills was dual to a topological string theory in
twistor space.  This led Cachazo, Svrcek and Witten
\cite{Cachazo:2004kj} to the conjecture that (as will be elaborated on
in section \ref{sec:csw}) tree-level amplitudes of gluons may be
calulated by sewing together certain on-shell amplitudes using scalar
propagators.  This procedure has since been known as the CSW rules.

The CSW conjecture has led to several developments in tree-level
calculations
\cite{Georgiou:2004wu,Georgiou:2004by,Wu:2004fb,Wu:2004jx,Su:2004ym,
  Bena:2004ry,Kosower:2004yz,Giombi:2004ix,
  Dixon:2004za,Bern:2004ba,Badger:2004ty,
  Birthwright:2005ak,Birthwright:2005vi,Marquard:2005rh,Schwinn:2005pi},
and it was also used for some one-loop calculations in an approach
developed by Brandhuber, Spence and Travaglini
\cite{Brandhuber:2004yw, Quigley:2004pw, Bedford:2004py,
  Bedford:2004nh}.  It was somewhat unexpected that the rules did
apply to loop calculations, but the discrepancy was resolved by the
introduction of the 'holomorphic anomaly' \cite{Cachazo:2004by}. When
combined with the unitarity method of Bern, Dixon, Dunbar and Kosower
\cite{Bern:1994zx,Bern:1994cg,Bern:2004bt} the concept led to a method
for calculating one-loop amplitudes in gauge theory as well as to
extensive studies of the twistor space structure of one-loop
amplitudes \cite{Cachazo:2004dr,Britto:2004nj,Bidder:2004tx,
  Britto:2004tx,Bidder:2004vx,Bern:2004bt}.

Inspired by this, but without depending on it, Britto, Cachazo and
Feng showed that, through the concept of generalized unitarity, the
calculation of one-loop amplitudes in $\N=4$ Super Yang--Mills could
be reduced to the calculation of certain on-shell tree amplitudes
\cite{Britto:2004nc}. Some applications are
\cite{Bidder:2005ri,Brandhuber:2005jw,Buchbinder:2005wp,Risager:2005ke}.
When using infrared concistency conditions on the one-loop amplitudes
it is possible to obtain compact expressions for tree amplitudes
\cite{Bern:2004ky, Bern:2004bt,Roiban:2004ix}, and this was used by
the authors mentioned above to develop recursion relations
\cite{Britto:2004ap} for tree amplitudes. A proof of the latter was
developed together with Witten
\cite{Britto:2005fq}, giving rise to the term BCFW recursion. This is
elaborated on in section \ref{sec:csw}.

The methods of BCFW recursion have been applied and extended with
great success to both tree-level calculations in Yang--Mills and other
theories \cite{Luo:2005rx,
  Luo:2005my,Bedford:2005yy,Cachazo:2005ca,Britto:2005dg,Badger:2005zh,
  Badger:2005jv,Forde:2005ue} as well as one-loop calculations
\cite{Bern:2005hs,Bern:2005ji, Bern:2005cq,Bern:2005hh}, and, in some
of these cases, been merged with technology such as unitarity methods
(see above).

The development described here has taken place alongside a
development in string theory and these two have interacted and
inspired each other. The string theory development and its interaction
with the one described here, is reviewed in \cite{Cachazo:2005ga}.

The purpose of the present letter is to prove the CSW rules for gluons
in external states using a method closely related to BCFW recursion.
This will tie together the two main developments in tree-level
calculations described here and show that one follows from the other.
In addition, it may shed some light on possible generalizations of the
CSW rules.

The contents of this paper are as follows: In section \ref{sec:not} we
review our notation and conventions. In sections \ref{sec:csw} and
\ref{sec:bcfw} we review the CSW and BCFW methods before proceding to
the proof of the CSW rules in section \ref{sec:proof}. In section
\ref{sec:dis} the implications of the proof are discussed.

\section{Notation and Conventions}
\label{sec:not}

We will be using the notion of colour ordering and the spinor helicity
notation. These matters are very well described in \cite{Dixon:1996wi}
whose conventions we use. For the spinor helicity notation, we use the
shorthand notation
\begin{equation}
\la j^+|\slash\hskip -8pt K|i^+\ra=  
\la i^-|\slash\hskip -8pt K|j^-\ra = \la iKj],
\end{equation}
\begin{equation}
  \la i^-|=\la i|,\quad \la i^+|=[i|,\quad
|i^+\ra=|i\ra,\quad |i^-\ra=|i].
\end{equation}
The derivations all rely on the underlying space being complexified
Minkowski space. This means that we may choose the two spinors $|i\ra$
and $|i]$ associated with $k_i^\mu$ to be complex and independent. By
convention, $|i\ra$ is referred to as the holomorphic spinor and $|i]$
is known as the anti-holomorphic spinor.

\section{The CSW Rules}
\label{sec:csw}

The starting point of the CSW rules is the amplitudes
\cite{Parke:1986gb,Berends:1987me},
\begin{eqnarray}
  A(1^\pm,2^+,3^+,\ldots,n^+)&=&0,\label{eq:lmhv}\\
A(1^+,2^+,\ldots,r^-,\ldots,s^-,\ldots,n^+)&=&
i\frac{\la rs\ra^4}{\la 12\ra\la 23\ra\cdots\la n1\ra}.
\label{eq:parketaylor}
\end{eqnarray}
The latter is a 'maximally helicity violating' (MHV) amplitude
whose main characteristic, in this context, is that it depends only on
the holomorphic spinors of the incoming particles. 

The CSW rules state how an amplitude can be calculated as the sum of
contributions each represented by a diagram similar to a Feynman
diagram. The vertices of such a diagram have two negative helicity
gluons and any number of positive helicity gluons, and the edges
connect gluons of negative and positive helicity. The contribution of
such a diagram is given as the product of the on-shell amplitudes
(\ref{eq:parketaylor}) for each MHV vertex and the product of usual
scalar propagators for each edge. The entire amplitude is the sum of
all possible CSW diagrams. Any CSW diagram contibuting to an amplitude
with $n$ negative helicity gluons (an N$^{n-2}$MHV amplitude, N for
'next-to-') contains $n-2$ propagators, and every diagram is exactly
characterized by its propagators.

The internal lines are not lightlike, and thus have no \emph{a priori}
spinors to put into (\ref{eq:parketaylor}) at each vertex. Instead,
one choses a lightlike momentum $\eta^\mu$ and writes the internal
momentum, say $P_i$, as
\begin{equation}
  \label{eq:projection}
  P_i^\mu=P_i^{\flat\mu}+z\eta^\mu,\qquad z=\frac{P_i^2}{2P_i\!\cdot\!\eta},
\end{equation}
such that $P_i^{\flat\mu}$ is lightlike. It is the spinors of this
momentum which are used for the vertices. Because the expressions
always turn out invariant under scaling of the internal spinors, we may use
$|P_i\eta]$ as the holomorphic spinor and $|P_i\eta\ra$ as the
anti-holomorphic spinor. Notice, however, that the latter is never
used in the CSW rules, and the end expression is independent of
$|\eta\ra$. Notice also, that even though we choose a displaced
version of the internal momenta for defining the vertices, the momenta
appearing in the propagators are undisplaced.

\section{BCFW Recursion}
\label{sec:bcfw}

BCFW on-shell recursion is a way to reconstruct a scattering amplitude
from its singularities. In its pure form, it consists of adding a
momentum $z\eta^\mu$ on one external particle and subtracting it on
another while choosing $\eta^\mu$ such that this shift of the two
momenta leaves them on-shell. $z$ is a complex variable. This defines
a shifted amplitude $\widehat A(z)$ where we would like to know
$\widehat A(0)$. Provided that $\widehat A(z)\to 0$ as $z\to\infty$ we
can write
\begin{eqnarray}
  0&=&\frac1{2\pi i}\oint_{{\mathcal C}{\rm ~at~}\infty}\frac{\widehat A(z)}z
=\widehat A(0)-\sum_{{\rm poles~}z_i}\frac{{\rm Res}_{z_i}
\widehat A(z)}{z_i},
\label{eq:contour}
\end{eqnarray}
By inspecting the Feynman rules for a shifted amplitude, we see that
there are two sources of poles in the shifted amplitude. The one is
the external polarizations where the spinor of an external particle
appears in the denominator. This type of pole must be removable since
choosing $\eta$ itself as the reference momentum for the polarization
removes the $z$ dependence of the propagator. We may still make an
other choice of polarization reference momentum and use that to lower
the power of asymptotic $z$ dependence of the shifted amplitude, but
we need not consider the resultant pole in (\ref{eq:contour}). The
other source is the propagators. There may be cases where the
singularity coming from a propagator is not physical (\emph{i.e.},
removable) but in that case the residue is zero and it makes no
difference whether we include the contribution or not. In conclusion,
the poles in $z$ we need to consider are those that arise from
propagators which are affected by the shift.
It is easily seen that the residue at such a pole is proportional
to the product of the on-shell amplitudes at each end of the
propagator in question, or
\begin{equation}
\label{eq:bcfw1}
  A(0)=\sum_{i\in \P}\widehat
  A^{L,i}\left(-\frac{P_i^2}{2P_i\!\cdot\!\eta}\right)
\frac{i}{P_i^2}\widehat A^{R,i}\left(-\frac{P_i^2}{2P_i\!\cdot\!\eta}\right),
\end{equation}
where $\P$ is the set of shifted propagators and $\widehat
A^{L,i}(z)$, $\widehat A^{R,i}(z)$ are the two shifted amplitudes
separated by $P_i$. In this expression, the particles entering and
leaving the propagator must be of opposite helicity. We choose to
regard two propagators with opposite helicity assignment as distinct
although they have the same momentum running through them. Since they
never appear in the same diagram, this will not cause problems for the
arguments that follow.

The above discussion can be made concrete by a rederivation of
(\ref{eq:parketaylor}).  Consider the amplitude
$A(1^-,2^+,\ldots,m^-,\ldots,n^+)$ and choose the reference momentum
to be $|1\ra[n|$ such that
\begin{eqnarray}
  \hat{|1\ra}\hat{[1|}&=&|1\ra\big([1|+z[n|\big),\\
  \hat{|n\ra}\hat{[n|}&=&\big(|n\ra-z|1\ra\big)[n|.
\end{eqnarray}
In other words, $|1\ra$ and $|n]$ are unchanged. As $z\to\infty$, the
shifted amplitude goes as $z^{-1}$ because the worst Feynman diagram,
containing only three-vertices, contributes $z$ from vertices and
propagators and $z^{-1}$ from each of the polarizations. Most terms in
the sum (\ref{eq:bcfw1}) are in fact zero since one of the shifted
amplitudes always falls in the category of (\ref{eq:lmhv}). That
equation, however, does not hold exactly for complex momenta, as the
amplitude $A(1^+,2^+,3^-)$ is only zero if one of the square brackets
between two of the three particles is zero. The two diagrams which
contain this so-called googly 3-vertex are
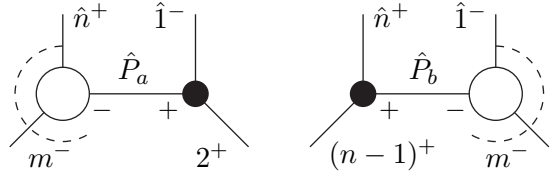
\begin{center}
  \begin{picture}(110,70)
\Line(40,30)(80,30)
\Line(80,30)(80,60)
\Line(80,30)(100,10)
\Line(30,40)(30,60)
\Line(23,23)(10,10)
\Vertex(80,30){5}
\Text(77,60)[r]{$\hat 1^-$}
\Text(34,60)[l]{$\hat n^+$}
\Text(57,40)[c]{$\hat P_a$}
\Text(70,24)[c]{\small $+$}
\Text(45,24)[c]{\small $-$}
\Text(17,7)[l]{$m^-$}
\Text(93,7)[r]{$2^+$}
\CArc(30,30)(10,0,360)
\DashCArc(30,30)(18,90,300){3}
  \end{picture}
  \begin{picture}(110,70)
\Line(30,30)(70,30)
\Line(80,40)(80,60)
\Line(87,23)(100,10)
\Line(30,30)(30,60)
\Line(30,30)(10,10)
\Vertex(30,30){5}
\Text(77,60)[r]{$\hat 1^-$}
\Text(34,60)[l]{$\hat n^+$}
\Text(53,40)[c]{$\hat P_b$}
\Text(65,24)[c]{\small $-$}
\Text(40,24)[c]{\small $+$}
\Text(17,7)[l]{$(n-1)^+$}
\Text(93,7)[r]{$m^-$}
\CArc(80,30)(10,0,360)
\DashCArc(80,30)(18,240,90){3}
  \end{picture}
\end{center}
In the first of these, $\hat P_a^2=(\hat p_1+p_2)^2=[\hat 12] \la
2\hat 1\ra=0$ requires either $[\hat 12]$ or $\la 21\ra$ to be 0.
Since $\la 21\ra$ is externally defined, it must be $[\hat 12]$, and
hence, the googly amplitude vanishes. In the second diagram an
identical argument means that $\la\hat n(n-1)\ra$ must vanish. 
That does not interfere with the 3-point googly vertex, but the 3-point
MHV would be zero in that case (as it is for real momenta). Notice,
that the shifted 3-point googly disappears when the shift is made
in the anti-holomorphic spinor and the shifted 3-point MHV disappears
when the shift is made in the holomorphic spinor.

Assuming that
(\ref{eq:parketaylor}) applies to an amplitude with $n-1$ external
gluons, we get for the amplitude in question
\begin{eqnarray}
  A(0)&=&\frac{i\la 1m\ra^4}{\la \hat P_b 1\ra\la 12\ra\cdots
\la (n-2)\hat P_b\ra}\frac{i}{P_b^2}\frac{i[n(n-1)]^4}
{[\hat P_b(n-1)][(n-1)n][n\hat P_b]}\nonumber\\
&=&i\frac{\la 1m\ra^4}{\la 12\ra\cdots\la (n-3)(n-2)\ra}
\frac{[n(n-1)]^3}{\la 1\hat P_b(n-1)]\la (n-2)\hat P_bn]}
\frac1{[n(n-1)]\la (n-1)n\ra}\nonumber\\
&=&i\frac{\la 1m\ra^4}{\la 12\ra\cdots\la (n-3)(n-2)\ra\la (n-1)n\ra}
\frac{[(n-1)n]^2}{\la 1n\ra[n(n-1)]\la (n-2)(n-1)\ra[(n-1)n]}\nonumber\\
&=&i\frac{\la 1m\ra^4}{\la 12\ra\cdots\la n1\ra}.
\end{eqnarray}
This provides a proof by induction of (\ref{eq:parketaylor}). In this
derivation we never needed to use the actual value of $z_i$, something
which we will need in general.

Before proceding to the direct proof of the CSW rules, it should be
noted that BCFW recursion can reconstruct any gluon amplitude solely
from the knowledge of its singularities. Since the CSW rules are known
to provide results which are Lorentz and gauge invariant and which
have the right singularities, the existence of BCFW recursion provides
an indirect proof of the CSW rules, as was noted in
\cite{Britto:2005fq}.

\section{Proof}
\label{sec:proof}

We prove the CSW rules by induction. First, we choose a shift of the
external momenta which will allow us to prove that an NMHV amplitude
may be calculated by means of the CSW rules. We then employ a
generalized version of that shift to a N$^{n-1}$MHV amplitude in order
to show that we may calculate it by the CSW rules provided that we can
calculate all more helicity violating amplitudes by the CSW rules.

Instead of choosing shifts which minimize the terms to sum over in
(\ref{eq:bcfw1}), we choose shifts that affect every propagator which
may occur in a CSW diagram. A propagator is characterized by a
consecutive set of external particles ($i,\ldots,j$) whose total
momentum runs through it, and the propagator occurs in a CSW diagram
if (and only if) the set and its compliment include at least one gluon
of negative helicity each.  Exactly this set of propagators is
affected by a shift on every negative helicity gluon, as long as the
sum of any subset of the shifts does not vanish. In addition, if all
3-point googly amplitudes are to drop out, the shifts must all involve
the anti-holomorphic spinors as was seen in the previous section.

For the NMHV amplitude, these arguments suggest the choice,
\begin{eqnarray}
  |\hat m_1\ra[\hat m_1|&=&
|m_1\ra([m_1|+z\la m_2m_3\ra[\eta|),\nonumber\\
  |\hat m_2\ra[\hat m_2|&=&
|m_2\ra([m_2|+z\la m_3m_1\ra[\eta|),\\
  |\hat m_3\ra[\hat m_3|&=&
|m_3\ra([m_3|+z\la m_1m_2\ra[\eta|),\nonumber
\end{eqnarray}
where $m_{1,2,3}$ are the three negative helicity gluons. The sum of
the momentum shifts vanishes by the Schouten identity. As $z\to
\infty$, the most dangerous Feynman diagram, composed of
three-vertices only, goes as $z^{-2}$ because the vertices and
propagators contribute a $z$ while the polarizations contribute
$z^{-3}$. Thus, all conditions for performing BCFW recursion with this
shift are satisfied.

The result of this is evidently the sum of all propagators which may
appear in the CSW diagram times two MHV amplitudes, one for each end
of the propagator, both of which are shifted in such a way that the
propagator goes on-shell,
\begin{equation}
  \widehat A_3(0)=\sum_{i\in\P_3}\widehat A_2^{1,i}\left(
-\frac{P_i^2}{2\la\lambda_iP_i\eta]}\right)\frac{i}{P_i^2}
\widehat A_2^{2,i}\left(-\frac{P_i^2}{2\la\lambda_iP_i\eta]}\right).
\end{equation}
In this expression $\la\lambda_i|$ is short for the spinor multiplying
$z[\eta|$ in the shifted propagator momentum; in the NMHV case, all
$\la\lambda_i|$ are equal to $\la m_1m_2\ra\la m_3|$ or one of its
cyclic permutations. The subscripts on the $A$'s denote the number of
negative helicity gluons.

The shifted MHV amplitudes of the expression are completely
holomorphic, and since the shift is purely anti-holomorphic, they are
not influenced. The momentum running in the propagator $i$ has been
put on-shell using the reference momentum $|\lambda_i\ra[\eta|$, and
it takes part in the MHV vertices in the same manner as in the CSW
rules.  Hence its holomorohic spinor can be written as $|P_i\eta]$ as
in those rules. By collecting all these facts, it is seen that an
NMHV amplitude may indeed be calculated by the CSW rules.

Let us now assume that the CSW rules hold for any N$^p$MHV amplitude
when $p<n-1$ and consider an N$^{n-1}$MHV amplitude. We now shift the
momenta of all the $n+1$ negative helicity gluons as
\begin{equation}
  |\hat m_i\ra[\hat m_i|
=|m_i\ra([m_i|+zr_i[\eta|),
\end{equation}
where $\sum_{i=1}^{n+1}|m_i\ra r_i=0$ without the same holding for any
subset of them. By the same argument as for NMHV, the shifted
amplitude goes as $z^{-n}$ as $z\to\infty$. The $z$-dependent
propagators are exactly those that appear in the CSW rules, and the
$z$ integration is frozen on each of these in turn,
\begin{equation}
\label{eq:decom}
  A_{n+1}(0)=\sum_{i\in\P_{n+1}}\widehat
  A_{m(i)}^{L,i}\left(-\frac{P_i^2}{\la \lambda_i 
P_i\eta]}\right)\frac{i}{P_i^2}\widehat A_{n+2-m(i)}^{R,i}
\left(-\frac{P_i^2}{\la \lambda_i P_i\eta]}\right). 
\end{equation}
Again $|\lambda_i\ra$ denotes the sum of $|m_j\ra r_j$ over the
momentum shifts isolated on one side of the propagator. In this
expression, the $\widehat A$'s have a lower number of negative
helicity gluons than the original amplitude.

Because of our induction assumption we may calculate the $\widehat
A$'s using the CSW rules, in which case (\ref{eq:decom}) becomes a sum
of terms each consisting of $n$ MHV vertices, one unshifted
propagator, and $n-2$ propagators shifted as to put the unshifted one
on-shell. In the same way we argued above, we may take $|P_i\eta]$ to
be the holomorphic spinor of $\Phat_i$, and as above, the MHV vertices
are independent of the shift. This leads us to conclude that every
term in the above-mentioned expansion of (\ref{eq:decom}) represents a
CSW diagram for the N$^{n-1}$MHV amplitude where the MHV vertices are
equal to those given by the CSW rules while all but one of the
propagators are shifted.

Every N$^{n-1}$MHV CSW diagram appears many times in this way but with
different propagators unshifted. Let us consider one N$^{n-1}$MHV CSW
diagram and extract all terms in (\ref{eq:decom}) where it appears.
Since the sum runs over all possible CSW propagators, we obtain
contributions exactly from the part of the sum which runs over the
$n-1$ propagators of the CSW diagram in question. For each of these,
there are unique CSW decompositions of the two shifted amplitudes
which yield the set of propagators we seek. The contributions to the
diagram from (\ref{eq:decom}) thus sums up to the MHV vertices
prescribed by the CSW rules times
\begin{equation}
\label{eq:shifsum}
  \sum_{i=1}^{n-1}\frac{i}{P_i^2}\prod_{j=1,j\neq i}^{n-1}
\frac{i}{\widehat P^2_{ji}},
\end{equation}
where 
\begin{equation}
  \widehat P_{ji}^2=(P_j+z_i|\lambda_j\ra[\eta|)^2
=P_j^2-P_i^2\frac{\la\lambda_jP_j\eta]}{\la\lambda_iP_i\eta]},
\end{equation}
is propagator $j$ when its momentum is shifted such that propagator
$i$ goes on-shell. Equation (\ref{eq:shifsum}) however, is just the product of
unshifted propagators. This can be realized by making the shift
\begin{equation}
  \prod_{i=1}^{n-1}\frac{i}{P_i^2}\to
 \prod_{i=1}^{n-1}\frac{i}{(P_i+z|\lambda_i\ra[\eta|)^2},
\end{equation}
and using (\ref{eq:contour}) in the same way as to derive
(\ref{eq:bcfw1}). Thus, the contribution from the CSW diagram
considered is exactly as given by the CSW rules.

We can apply this argument to any N$^{n-1}$MHV CSW diagram. Since we
have also shown that every part of (\ref{eq:decom}) can be written in
a way where it contributes to an N$^{n-1}$MHV CSW diagram as described,
it follows that the N$^{n-1}$MHV CSW rules are in one-to-one
correspondence with (\ref{eq:decom}). By induction, any amplitude can
thus be calculated by the CSW rules.

\section{Discussion}
\label{sec:dis}

As noted at the end of section \ref{sec:bcfw}, the CSW rules have been
known to give the right results, but only by an indirect proof. The
present proof provides an explanation of why the rules hold true, and
what is their connection to BCFW recursion.

The proof also clarifies a point in the formulation of the CSW rules.
When constructing the internal spinors through (\ref{eq:projection}),
$\eta^\mu$ is assumed to be constant throughout the calculation, even
though this is only strictly necessary for the anti-holomorphic
spinor. The holomorphic spinor $|\eta\ra$ may differ significantly
from propagator to propagator.

Although this fact does not alter tree-level Yang--Mills calculations,
it may have implications in extensions of the CSW rules. When trying
to develop a CSW approach to a field theory where (the equivalent of)
MHV amplitudes are not holomorphic (\emph{e.g.}, gravity) there will
be a dependence on the holomorphic reference spinor which might not be
assumed to be constant. Also, when the CSW rules are applied to loop
calculations, the full reference momentum takes part in the
expressions, and a precise knowledgement of the possible holomorphic
reference spinors is needed. The fact that the MHV one-loop
calculations \emph{do} carry through in a formulation where the
holomorphic reference spinor is unconstrained
\cite{Brandhuber:2004yw,Quigley:2004pw,Bedford:2004py,Bedford:2004nh},
shows that there is probably more to the CSW rules than suggested by
this proof.

\begin{acknowledgments}
  I wish to thank Emil Bjerrum-Bohr, David Dunbar, Harald Ita, Warren
  Perkins and Poul Henrik Damgaard for their constructive comments and
  suggestions, and the Physics Department at Swansea University for
  its hospitality during some of my work. 
\end{acknowledgments}


\begin{thebibliography}{10}

\bibitem{Witten:2003nn}
E.~Witten, {\it Perturbative gauge theory as a string theory in twistor space},
   {\em Commun. Math. Phys.} {\bf 252} (2004) 189--258,
  [\href{http://xxx.lanl.gov/abs/hep-th/0312171}{{\tt hep-th/0312171}}].

\bibitem{Nair:1988bq}
V.~P. Nair, {\it A current algebra for some gauge theory amplitudes},  {\em
  Phys. Lett.} {\bf B214} (1988) 215.

\bibitem{Cachazo:2004kj}
F.~Cachazo, P.~Svrcek, and E.~Witten, {\it Mhv vertices and tree amplitudes in
  gauge theory},  {\em JHEP} {\bf 09} (2004) 006,
  [\href{http://xxx.lanl.gov/abs/hep-th/0403047}{{\tt hep-th/0403047}}].

\bibitem{Georgiou:2004wu}
G.~Georgiou and V.~V. Khoze, {\it Tree amplitudes in gauge theory as scalar mhv
  diagrams},  {\em JHEP} {\bf 05} (2004) 070,
  [\href{http://xxx.lanl.gov/abs/hep-th/0404072}{{\tt hep-th/0404072}}].

\bibitem{Georgiou:2004by}
G.~Georgiou, E.~W.~N. Glover, and V.~V. Khoze, {\it Non-mhv tree amplitudes in
  gauge theory},  {\em JHEP} {\bf 07} (2004) 048,
  [\href{http://xxx.lanl.gov/abs/hep-th/0407027}{{\tt hep-th/0407027}}].

\bibitem{Wu:2004fb}
J.-B. Wu and C.-J. Zhu, {\it Mhv vertices and scattering amplitudes in gauge
  theory},  {\em JHEP} {\bf 07} (2004) 032,
  [\href{http://xxx.lanl.gov/abs/hep-th/0406085}{{\tt hep-th/0406085}}].

\bibitem{Wu:2004jx}
J.-B. Wu and C.-J. Zhu, {\it Mhv vertices and fermionic scattering amplitudes
  in gauge theory with quarks and gluinos},  {\em JHEP} {\bf 09} (2004) 063,
  [\href{http://xxx.lanl.gov/abs/hep-th/0406146}{{\tt hep-th/0406146}}].

\bibitem{Su:2004ym}
X.~Su and J.-B. Wu, {\it Six-quark amplitudes from fermionic mhv vertices},
  {\em Mod. Phys. Lett.} {\bf A20} (2005) 1065--1076,
  [\href{http://xxx.lanl.gov/abs/hep-th/0409228}{{\tt hep-th/0409228}}].

\bibitem{Bena:2004ry}
I.~Bena, Z.~Bern, and D.~A. Kosower, {\it Twistor-space recursive formulation
  of gauge theory amplitudes},  {\em Phys. Rev.} {\bf D71} (2005) 045008,
  [\href{http://xxx.lanl.gov/abs/hep-th/0406133}{{\tt hep-th/0406133}}].

\bibitem{Kosower:2004yz}
D.~A. Kosower, {\it Next-to-maximal helicity violating amplitudes in gauge
  theory},  {\em Phys. Rev.} {\bf D71} (2005) 045007,
  [\href{http://xxx.lanl.gov/abs/hep-th/0406175}{{\tt hep-th/0406175}}].

\bibitem{Giombi:2004ix}
S.~Giombi, R.~Ricci, D.~Robles-Llana, and D.~Trancanelli, {\it A note on
  twistor gravity amplitudes},  {\em JHEP} {\bf 07} (2004) 059,
  [\href{http://xxx.lanl.gov/abs/hep-th/0405086}{{\tt hep-th/0405086}}].

\bibitem{Dixon:2004za}
L.~J. Dixon, E.~W.~N. Glover, and V.~V. Khoze, {\it Mhv rules for higgs plus
  multi-gluon amplitudes},  {\em JHEP} {\bf 12} (2004) 015,
  [\href{http://xxx.lanl.gov/abs/hep-th/0411092}{{\tt hep-th/0411092}}].

\bibitem{Bern:2004ba}
Z.~Bern, D.~Forde, D.~A. Kosower, and P.~Mastrolia, {\it Twistor-inspired
  construction of electroweak vector boson currents},  {\em Phys. Rev.} {\bf
  D72} (2005) 025006, [\href{http://xxx.lanl.gov/abs/hep-ph/0412167}{{\tt
  hep-ph/0412167}}].

\bibitem{Badger:2004ty}
S.~D. Badger, E.~W.~N. Glover, and V.~V. Khoze, {\it Mhv rules for higgs plus
  multi-parton amplitudes},  {\em JHEP} {\bf 03} (2005) 023,
  [\href{http://xxx.lanl.gov/abs/hep-th/0412275}{{\tt hep-th/0412275}}].

\bibitem{Birthwright:2005ak}
T.~G. Birthwright, E.~W.~N. Glover, V.~V. Khoze, and P.~Marquard, {\it
  Multi-gluon collinear limits from mhv diagrams},  {\em JHEP} {\bf 05} (2005)
  013, [\href{http://xxx.lanl.gov/abs/hep-ph/0503063}{{\tt hep-ph/0503063}}].

\bibitem{Birthwright:2005vi}
T.~G. Birthwright, E.~W.~N. Glover, V.~V. Khoze, and P.~Marquard, {\it
  Collinear limits in qcd from mhv rules},  {\em JHEP} {\bf 07} (2005) 068,
  [\href{http://xxx.lanl.gov/abs/hep-ph/0505219}{{\tt hep-ph/0505219}}].

\bibitem{Marquard:2005rh}
P.~Marquard and T.~G. Birthwright, {\it Multi gluon collinear limits from mhv
  amplitudes},  \href{http://xxx.lanl.gov/abs/hep-ph/0505264}{{\tt
  hep-ph/0505264}}.

\bibitem{Schwinn:2005pi}
C.~Schwinn and S.~Weinzierl, {\it Scalar diagrammatic rules for born amplitudes
  in qcd},  {\em JHEP} {\bf 05} (2005) 006,
  [\href{http://xxx.lanl.gov/abs/hep-th/0503015}{{\tt hep-th/0503015}}].

\bibitem{Brandhuber:2004yw}
A.~Brandhuber, B.~Spence, and G.~Travaglini, {\it One-loop gauge theory
  amplitudes in n = 4 super yang-mills from mhv vertices},  {\em Nucl. Phys.}
  {\bf B706} (2005) 150--180,
  [\href{http://xxx.lanl.gov/abs/hep-th/0407214}{{\tt hep-th/0407214}}].

\bibitem{Quigley:2004pw}
C.~Quigley and M.~Rozali, {\it One-loop mhv amplitudes in supersymmetric gauge
  theories},  {\em JHEP} {\bf 01} (2005) 053,
  [\href{http://xxx.lanl.gov/abs/hep-th/0410278}{{\tt hep-th/0410278}}].

\bibitem{Bedford:2004py}
J.~Bedford, A.~Brandhuber, B.~Spence, and G.~Travaglini, {\it A twistor
  approach to one-loop amplitudes in n = 1 supersymmetric yang-mills theory},
  {\em Nucl. Phys.} {\bf B706} (2005) 100--126,
  [\href{http://xxx.lanl.gov/abs/hep-th/0410280}{{\tt hep-th/0410280}}].

\bibitem{Bedford:2004nh}
J.~Bedford, A.~Brandhuber, B.~Spence, and G.~Travaglini, {\it
  Non-supersymmetric loop amplitudes and mhv vertices},  {\em Nucl. Phys.} {\bf
  B712} (2005) 59--85, [\href{http://xxx.lanl.gov/abs/hep-th/0412108}{{\tt
  hep-th/0412108}}].

\bibitem{Cachazo:2004by}
F.~Cachazo, P.~Svrcek, and E.~Witten, {\it Gauge theory amplitudes in twistor
  space and holomorphic anomaly},  {\em JHEP} {\bf 10} (2004) 077,
  [\href{http://xxx.lanl.gov/abs/hep-th/0409245}{{\tt hep-th/0409245}}].

\bibitem{Bern:1994zx}
Z.~Bern, L.~J. Dixon, D.~C. Dunbar, and D.~A. Kosower, {\it One loop n point
  gauge theory amplitudes, unitarity and collinear limits},  {\em Nucl. Phys.}
  {\bf B425} (1994) 217--260,
  [\href{http://xxx.lanl.gov/abs/hep-ph/9403226}{{\tt hep-ph/9403226}}].

\bibitem{Bern:1994cg}
Z.~Bern, L.~J. Dixon, D.~C. Dunbar, and D.~A. Kosower, {\it Fusing gauge theory
  tree amplitudes into loop amplitudes},  {\em Nucl. Phys.} {\bf B435} (1995)
  59--101, [\href{http://xxx.lanl.gov/abs/hep-ph/9409265}{{\tt
  hep-ph/9409265}}].

\bibitem{Bern:2004bt}
Z.~Bern, L.~J. Dixon, and D.~A. Kosower, {\it All next-to-maximally
  helicity-violating one-loop gluon amplitudes in n = 4 super-yang-mills
  theory},  {\em Phys. Rev.} {\bf D72} (2005) 045014,
  [\href{http://xxx.lanl.gov/abs/hep-th/0412210}{{\tt hep-th/0412210}}].

\bibitem{Cachazo:2004dr}
F.~Cachazo, {\it Holomorphic anomaly of unitarity cuts and one-loop gauge
  theory amplitudes},  \href{http://xxx.lanl.gov/abs/hep-th/0410077}{{\tt
  hep-th/0410077}}.

\bibitem{Britto:2004nj}
R.~Britto, F.~Cachazo, and B.~Feng, {\it Computing one-loop amplitudes from the
  holomorphic anomaly of unitarity cuts},  {\em Phys. Rev.} {\bf D71} (2005)
  025012, [\href{http://xxx.lanl.gov/abs/hep-th/0410179}{{\tt
  hep-th/0410179}}].

\bibitem{Bidder:2004tx}
S.~J. Bidder, N.~E.~J. Bjerrum-Bohr, L.~J. Dixon, and D.~C. Dunbar, {\it N = 1
  supersymmetric one-loop amplitudes and the holomorphic anomaly of unitarity
  cuts},  {\em Phys. Lett.} {\bf B606} (2005) 189--201,
  [\href{http://xxx.lanl.gov/abs/hep-th/0410296}{{\tt hep-th/0410296}}].

\bibitem{Britto:2004tx}
R.~Britto, F.~Cachazo, and B.~Feng, {\it Coplanarity in twistor space of n = 4
  next-to-mhv one-loop amplitude coefficients},  {\em Phys. Lett.} {\bf B611}
  (2005) 167--172, [\href{http://xxx.lanl.gov/abs/hep-th/0411107}{{\tt
  hep-th/0411107}}].

\bibitem{Bidder:2004vx}
S.~J. Bidder, N.~E.~J. Bjerrum-Bohr, D.~C. Dunbar, and W.~B. Perkins, {\it
  Twistor space structure of the box coefficients of n = 1 one-loop
  amplitudes},  {\em Phys. Lett.} {\bf B608} (2005) 151--163,
  [\href{http://xxx.lanl.gov/abs/hep-th/0412023}{{\tt hep-th/0412023}}].

\bibitem{Britto:2004nc}
R.~Britto, F.~Cachazo, and B.~Feng, {\it Generalized unitarity and one-loop
  amplitudes in n = 4 super-yang-mills},  {\em Nucl. Phys.} {\bf B725} (2005)
  [\href{http://xxx.lanl.gov/abs/hep-th/0412103}{{\tt hep-th/0412103}}].

\bibitem{Bidder:2005ri}
S.~J. Bidder, N.~E.~J. Bjerrum-Bohr, D.~C. Dunbar, and W.~B. Perkins, {\it
  One-loop gluon scattering amplitudes in theories with n < 4 supersymmetries},
   {\em Phys. Lett.} {\bf B612} (2005) 75--88,
  [\href{http://xxx.lanl.gov/abs/hep-th/0502028}{{\tt hep-th/0502028}}].

\bibitem{Brandhuber:2005jw}
A.~Brandhuber, S.~McNamara, B.~Spence, and G.~Travaglini, {\it Loop amplitudes
  in pure yang-mills from generalised unitarity},
  \href{http://xxx.lanl.gov/abs/hep-th/0506068}{{\tt hep-th/0506068}}.

\bibitem{Buchbinder:2005wp}
E.~I. Buchbinder and F.~Cachazo, {\it Two-loop amplitudes of gluons and
  octa-cuts in n = 4 super yang-mills},
  \href{http://xxx.lanl.gov/abs/hep-th/0506126}{{\tt hep-th/0506126}}.

\bibitem{Risager:2005ke}
K.~Risager, S.~J. Bidder, and W.~B. Perkins, {\it One-loop nmhv amplitudes
  involving gluinos and scalars in n = 4 gauge theory},
  \href{http://xxx.lanl.gov/abs/hep-th/0507170}{{\tt hep-th/0507170}}.

\bibitem{Bern:2004ky}
Z.~Bern, V.~Del~Duca, L.~J. Dixon, and D.~A. Kosower, {\it All
  non-maximally-helicity-violating one-loop seven-gluon amplitudes in n = 4
  super-yang-mills theory},  {\em Phys. Rev.} {\bf D71} (2005) 045006,
  [\href{http://xxx.lanl.gov/abs/hep-th/0410224}{{\tt hep-th/0410224}}].

\bibitem{Roiban:2004ix}
R.~Roiban, M.~Spradlin, and A.~Volovich, {\it Dissolving n = 4 loop amplitudes
  into qcd tree amplitudes},  {\em Phys. Rev. Lett.} {\bf 94} (2005) 102002,
  [\href{http://xxx.lanl.gov/abs/hep-th/0412265}{{\tt hep-th/0412265}}].

\bibitem{Britto:2004ap}
R.~Britto, F.~Cachazo, and B.~Feng, {\it New recursion relations for tree
  amplitudes of gluons},  {\em Nucl. Phys.} {\bf B715} (2005) 499--522,
  [\href{http://xxx.lanl.gov/abs/hep-th/0412308}{{\tt hep-th/0412308}}].

\bibitem{Britto:2005fq}
R.~Britto, F.~Cachazo, B.~Feng, and E.~Witten, {\it Direct proof of tree-level
  recursion relation in yang-mills theory},  {\em Phys. Rev. Lett.} {\bf 94}
  (2005) 181602, [\href{http://xxx.lanl.gov/abs/hep-th/0501052}{{\tt
  hep-th/0501052}}].

\bibitem{Luo:2005rx}
M.-x. Luo and C.-k. Wen, {\it Recursion relations for tree amplitudes in super
  gauge theories},  {\em JHEP} {\bf 03} (2005) 004,
  [\href{http://xxx.lanl.gov/abs/hep-th/0501121}{{\tt hep-th/0501121}}].

\bibitem{Luo:2005my}
M.-x. Luo and C.-k. Wen, {\it Compact formulas for all tree amplitudes of six
  partons},  {\em Phys. Rev.} {\bf D71} (2005) 091501,
  [\href{http://xxx.lanl.gov/abs/hep-th/0502009}{{\tt hep-th/0502009}}].

\bibitem{Bedford:2005yy}
J.~Bedford, A.~Brandhuber, B.~Spence, and G.~Travaglini, {\it A recursion
  relation for gravity amplitudes},  {\em Nucl. Phys.} {\bf B721} (2005)
  98--110, [\href{http://xxx.lanl.gov/abs/hep-th/0502146}{{\tt
  hep-th/0502146}}].

\bibitem{Cachazo:2005ca}
F.~Cachazo and P.~Svrcek, {\it Tree level recursion relations in general
  relativity},  \href{http://xxx.lanl.gov/abs/hep-th/0502160}{{\tt
  hep-th/0502160}}.

\bibitem{Britto:2005dg}
R.~Britto, B.~Feng, R.~Roiban, M.~Spradlin, and A.~Volovich, {\it All split
  helicity tree-level gluon amplitudes},  {\em Phys. Rev.} {\bf D71} (2005)
  105017, [\href{http://xxx.lanl.gov/abs/hep-th/0503198}{{\tt
  hep-th/0503198}}].

\bibitem{Badger:2005zh}
S.~D. Badger, E.~W.~N. Glover, V.~V. Khoze, and P.~Svrcek, {\it Recursion
  relations for gauge theory amplitudes with massive particles},  {\em JHEP}
  {\bf 07} (2005) 025, [\href{http://xxx.lanl.gov/abs/hep-th/0504159}{{\tt
  hep-th/0504159}}].

\bibitem{Badger:2005jv}
S.~D. Badger, E.~W.~N. Glover, and V.~V. Khoze, {\it Recursion relations for
  gauge theory amplitudes with massive vector bosons and fermions},
  \href{http://xxx.lanl.gov/abs/hep-th/0507161}{{\tt hep-th/0507161}}.

\bibitem{Forde:2005ue}
D.~Forde and D.~A. Kosower, {\it All-multiplicity amplitudes with massive
  scalars},  \href{http://xxx.lanl.gov/abs/hep-th/0507292}{{\tt
  hep-th/0507292}}.

\bibitem{Bern:2005hs}
Z.~Bern, L.~J. Dixon, and D.~A. Kosower, {\it On-shell recurrence relations for
  one-loop qcd amplitudes},  {\em Phys. Rev.} {\bf D71} (2005) 105013,
  [\href{http://xxx.lanl.gov/abs/hep-th/0501240}{{\tt hep-th/0501240}}].

\bibitem{Bern:2005ji}
Z.~Bern, L.~J. Dixon, and D.~A. Kosower, {\it The last of the finite loop
  amplitudes in qcd},  \href{http://xxx.lanl.gov/abs/hep-ph/0505055}{{\tt
  hep-ph/0505055}}.

\bibitem{Bern:2005cq}
Z.~Bern, L.~J. Dixon, and D.~A. Kosower, {\it Bootstrapping multi-parton loop
  amplitudes in qcd},  \href{http://xxx.lanl.gov/abs/hep-ph/0507005}{{\tt
  hep-ph/0507005}}.

\bibitem{Bern:2005hh}
Z.~Bern, N.~E.~J. Bjerrum-Bohr, D.~C. Dunbar, and H.~Ita, {\it Recursive
  calculation of one-loop qcd integral coefficients},
  \href{http://xxx.lanl.gov/abs/hep-ph/0507019}{{\tt hep-ph/0507019}}.

\bibitem{Cachazo:2005ga}
F.~Cachazo and P.~Svrcek, {\it Lectures on twistor strings and perturbative
  yang-mills theory},  {\em Proc. Sci.} {\bf RTN2005} (2005) 005,
  [\href{http://xxx.lanl.gov/abs/hep-th/0504194}{{\tt hep-th/0504194}}].

\bibitem{Dixon:1996wi}
L.~J. Dixon, {\it Calculating scattering amplitudes efficiently},
  \href{http://xxx.lanl.gov/abs/hep-ph/9601359}{{\tt hep-ph/9601359}}.

\bibitem{Parke:1986gb}
S.~J. Parke and T.~R. Taylor, {\it An amplitude for n gluon scattering},  {\em
  Phys. Rev. Lett.} {\bf 56} (1986) 2459.

\bibitem{Berends:1987me}
F.~A. Berends and W.~T. Giele, {\it Recursive calculations for processes with n
  gluons},  {\em Nucl. Phys.} {\bf B306} (1988) 759.

\end{thebibliography}

\providecommand{\href}[2]{#2}\begingroup\raggedright\endgroup

\end{document}